\documentclass{article}
\usepackage{spconf,amsmath,graphicx,hyperref}

\usepackage{float}
\usepackage{hyperref}
\hypersetup{
    colorlinks=true,
    linkcolor=blue,
    citecolor=blue,
    urlcolor=black
}
\usepackage{multicol, multirow}
\usepackage{tabularx}
\usepackage[table]{xcolor}
\usepackage{colortbl}

\usepackage{amsthm}
\usepackage{booktabs}
\usepackage{algorithm}
\usepackage{algorithmic}
\usepackage[switch]{lineno}

\usepackage{cite}
\usepackage{amssymb,amsfonts}

\usepackage{textcomp}

\usepackage[utf8]{inputenc} 
\usepackage[T1]{fontenc}    
\usepackage{hyperref}       
\usepackage{url}            
\usepackage{booktabs}       
\usepackage{amsfonts}       
\usepackage{nicefrac}       
\usepackage{microtype}      

\usepackage{enumitem}

\title{AUDIO DEEPFAKE DETECTION AT THE FIRST GREETING: ``HI!''}
\name{Haohan Shi$^1$, Xiyu Shi$^1$, Safak Dogan$^1$, Tianjin Huang$^2$, Yunxiao Zhang$^{2,*}$\thanks{This research was funded by Loughborough University (Grant No. GS1016) and the China Scholarship Council (Grant No. 202208060237). *Corresponding author.}}
\address{$^1$Institute for Digital Technologies, Loughborough University London, E20 3BS, UK\\
$^2$Department of Computer Science, University of Exeter, EX4 4QE, UK}
\begin{document}
%
\maketitle
\begin{abstract}
This paper focuses on audio deepfake detection under real-world communication degradations, with an emphasis on ultra-short inputs (0.5–2.0s), targeting the capability to detect synthetic speech at a conversation opening, e.g., when a scammer says “Hi.” We propose Short-MGAA (S-MGAA), a novel lightweight extension of Multi-Granularity Adaptive Time–Frequency Attention, designed to enhance discriminative representation learning for short, degraded inputs subjected to communication processing and perturbations. The S‑MGAA integrates two tailored modules: a Pixel–Channel Enhanced Module (PCEM) that amplifies fine‑grained time–frequency saliency, and a Frequency Compensation Enhanced Module (FCEM) to supplement limited temporal evidence via multi‑scale frequency modeling and adaptive frequency–temporal interaction. Extensive experiments demonstrate that S-MGAA consistently surpasses nine state-of-the-art baselines while achieving strong robustness to degradations and favorable efficiency–accuracy trade-offs, including low RTF, competitive GFLOPs, compact parameters, and reduced training cost, highlighting its strong potential for real-time deployment in communication systems and edge devices.
\end{abstract}

\section{Introduction}

The rapid advancement of deep generative models has expanded the production and spread of synthetic speech, raising significant concerns regarding deepfake audio misuse and its societal risks \cite{li2025survey}. 
In response, the field of Audio Deepfake Detection (ADD) has progressed rapidly, with competitions such as ASVspoof \cite{asvspoof2019,asvspoof2021,wang2024asvspoof} and ADD2022 \cite{add2022} advancing datasets, methods, and benchmarks.

In real-world communication scenarios, audio is often corrupted by codec compression and packet losses, which substantially degrade signal quality and reduce the robustness of ADD methods. 
While state-of-the-art (SOTA) methods perform well under clean, high-fidelity audio inputs, their robustness under real-world communication degradation remains limited.
Several studies have explored approaches to mitigate such degradations \cite{eusipcohaohan,nipshaohan}. However, most existing methods are designed under fixed audio input lengths (3–4s) \cite{eusipcohaohan,nipshaohan,aasist,xie2023domain,rawgat,li2022comparative}.
For shorter inputs, truncation or padding is commonly applied, leading to a duration mismatch between training and testing that degrades accuracy.
Artificially extending inputs at inference time is also impractical for latency-sensitive applications \cite{rosello2023conformer,lowrank}.

\begin{figure*}[ht]
\centering
\includegraphics[width=0.7\textwidth]{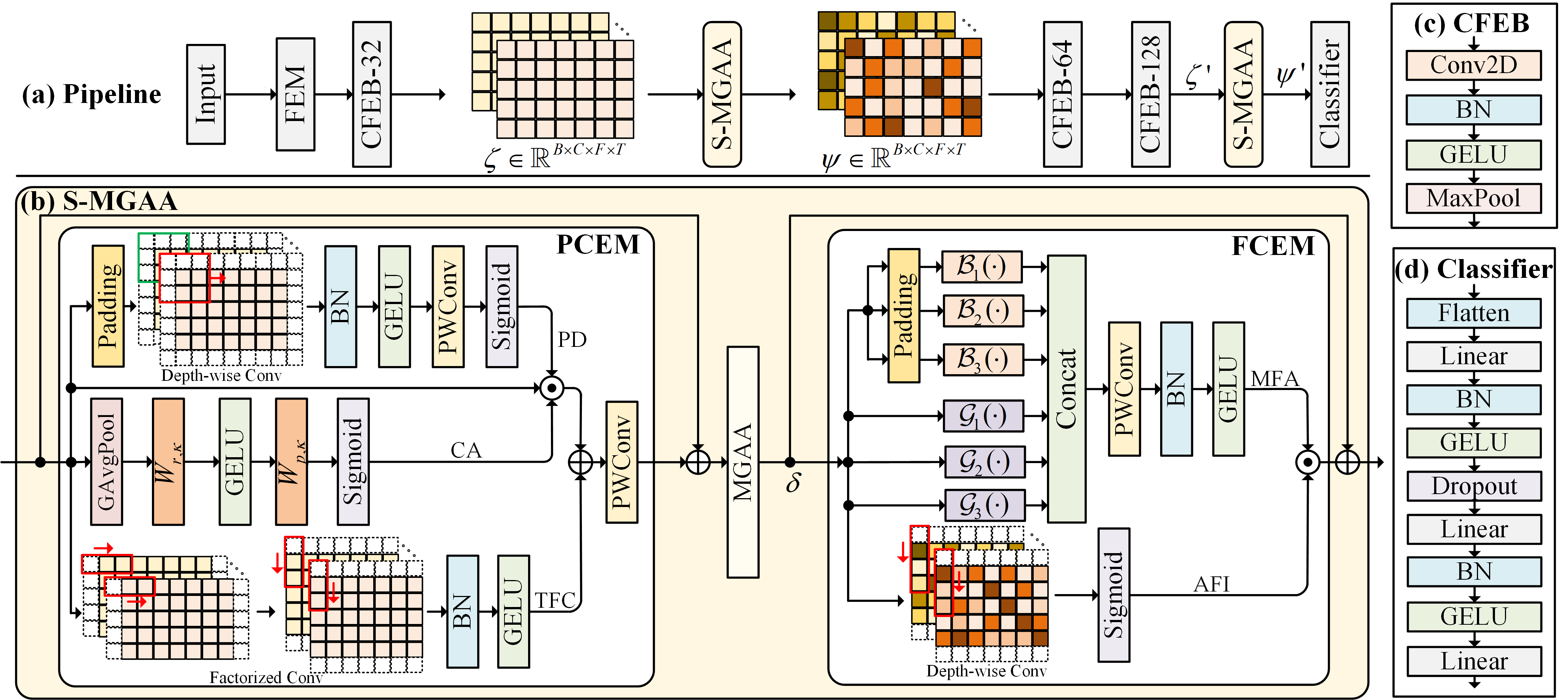}
\vspace{-0.4cm}
\caption{Proposed ADD framework for ultra-short-duration audio inputs. (a) The processing pipeline; (b) Short Multi-Granularity Adaptive Time-Frequency Attention; (c) Convolutional Feature Embedding Blocks; (d) The Classifier.}
\label{arch}
\vspace{-0.2cm}
\end{figure*}

Recently, short-duration ADD has drawn increasing attention. This work specifically targets the challenge of detecting deepfakes from ultra-short inputs (0.5–2.0s), allowing detection at the very onset of speech communication; for instance, when a scammer utters the initial ``Hi''. Such capability is critical for real-time security applications, where rapid detection can prevent social engineering attacks before the conversation progresses. 
To be practically deployable, the method must achieve robustness under real-world communication degradations and remain lightweight enough for deployment on resource-constrained devices such as smartphones.
However, most existing methods for short-duration ADD overlook both the robustness against real-world communication degradations and the lightweight design requirements \cite{zhang2024improving,xiao2025xlsr,10605999, zhang2022partialspoof}. 
The objective of this work is to address reliable ADD within the first seconds of speech conversation, ensuring both efficiency and robustness for real-world conversational settings.

To this end, we propose S-MGAA, a novel lightweight ADD framework tailored for ultra-short inputs under communication degradations. Our contributions are threefold:

\begin{itemize}[leftmargin=*]
\vspace{-0.22cm}
\item 
To the best of our knowledge, this is the first study that jointly ensures robustness to real-world degradations and considers ultra-short inputs (0.5–2.0s).
\vspace{-0.22cm}
\item We introduce a lightweight framework that enhances discriminative representation learning for degraded and ultra-short duration inputs while maintaining deployability in resource-constrained environments. 
\vspace{-0.22cm}
\item The proposed framework surpasses nine SOTA baselines on 0.5-2.0s inputs under diverse communication degradations, demonstrating a favorable trade-off among Real-Time Factor (RTF), training time, Giga Floating Point Operations Per Second (GFLOPs), and parameter count. These results demonstrate its capacity for real-time ADD within the opening seconds of a conversation in communication systems.
\end{itemize}

\section{Methodology}

\vspace{-0.2cm}
\subsection{Framework Architecture}
\vspace{-0.1cm}
Our work extends the Multi-Granularity Adaptive Time- Frequency Attention (MGAA) framework for ADD \cite{nipshaohan} to ultra‑short utterances (0.5s–2s). Although MGAA performs well on 4s clips under various real-world communication degradations, its accuracy drops on short inputs due to sparse, low‑saliency spoofing cues. 
To address this, we propose Short‑MGAA (S‑MGAA), an ultra‑short ADD architecture specifically designed to enhance representation learning for short-duration inputs, as shown in Fig. \ref{arch}.

The proposed S-MGAA is composed of three sequential modules: (1) a PCEM for saliency amplification at fine-grained Time-Frequency (TF) locations for both pixel and channel levels;
(2) the MGAA block; (3) a FCEM to supplement the limited temporal feature by exploiting multi-perspective frequency-domain features.
Three widely used TF features are selected for evaluation: Linear-Frequency Cepstral Coefficients (LFCC), Constant-Q Cepstral Coefficients (CQCC), and Mel-frequency Cepstral Coefficients (MFCC).
Denote the input to S-MGAA as \(\zeta\in \mathbb{R}^{B\times C\times F \times T}\), where \(B\) is the batch size, \(C\) the feature channel dimensions, \(F\) and \(T\) the frequency and temporal dimensions, respectively.
S-MGAA processes \(\zeta\) to generate \(\psi\in \mathbb{R}^{B\times C\times F\times T}\), where discriminative forgery-related cues are amplified and localized. \(\psi\) is further processed by CFEB-64 and CFEB-128 to obtain a higher-level representation \(\zeta'\). A second-stage S-MGAA is applied, resulting in the final output \(\psi'\), which is flattened and passed to the classifier for a binary decision.

\vspace{-0.2cm}
\subsection{Pixel-Channel Enhanced Module (PCEM)}
\vspace{-0.1cm}

PCEM aims to address the sparse and limited forgery TF features affected by ultra-short-duration audio and real-world communication degradations.
It systematically enhances the discriminative capacity of the model by jointly modeling pixel-level saliency, channel-wise importance, and TF coupling. The output of PCEM is computed as:
\begin{equation} 
\text{PCEM}(\zeta) = \mathcal{V}_{c}(\zeta \odot \mathcal{P}(\zeta) \odot \mathcal{C}(\zeta) + \mathcal{T}(\zeta)),
\end{equation}
where $\odot$ denotes element-wise multiplication, $\mathcal{V}_{c}(\cdot)$ the pointwise convolution for inter-channel mixing, and $\mathcal{P}(\cdot)$, $\mathcal{C}(\cdot)$, and $\mathcal{T}(\cdot)$ the Pixel-level Detector (PD), Channel-wise Amplifier (CA), and TF Coupling components (TFC), respectively. 
The PD highlights fine-grained pixel-level forgery clues and is computed as:
\begin{equation}
\mathcal{P}(\zeta) = \sigma_{s}(\mathcal{V}_{c}(\mathcal{\sigma}_{g}(\mathcal{N}_{b}(\mathcal{H}_{DW}(\zeta))))),
\end{equation}
where $\mathcal{H}_{DW}$ is a depthwise ($3,3$) convolution, $\mathcal{N}_{b}$ the batch normalization, $\sigma_{g}$ the Gaussian Error Linear Unit (GELU) and $\sigma_{s}$ the sigmoid activation. 
The CA emphasizes the importance of gathering clues across various channels and is computed as:
\begin{equation}
\mathcal{C}(\zeta) = \sigma_{s}(W_{p,\kappa}\cdot\sigma_{g}(W_{r,\kappa}\cdot\mathcal{G}_{avg}(\zeta))),
\end{equation}
where $\mathcal{G}_{avg}(\cdot)$ denotes global average pooling, $W_{r,\kappa}$ and $W_{p,\kappa}$ the channel compression and expansion convolution with ratio $\kappa=8$.
The TFC is computed by $\mathcal{T}(\zeta) = \sigma_{g}(\mathcal{N}_{b}(\mathcal{V}_{f}(\mathcal{V}_{t}(\zeta))))$, where $\mathcal{V}_{f}$ and $\mathcal{V}_{t}$ are factorized convolutions with kernel size $(1,3)$ and $(3,1)$, respectively, enabling joint modeling of frequency and temporal dependencies crucial for ultra-short-duration audio.

\begin{table*}[ht]
\caption{Detection performance of EER (\%) across input durations of 0.5s, 1s, 1.5s, and 2s. Experiments were repeated three times with different random seeds, and average metric values are reported. Bold entries indicate the best overall performance.}
\label{mainresult}
\centering
\setlength{\tabcolsep}{1.25mm}
\scalebox{0.6}{
\begin{tabular}{ccccccccccccccccccccccccccccc}
\toprule
\multirow{2}{*}{Model} & \multicolumn{7}{c}{0.5s} & \multicolumn{7}{c}{1.0s} & \multicolumn{7}{c}{1.5s} & \multicolumn{7}{c}{2.0s}\\ \cmidrule(r){2-8}  \cmidrule(r){9-15} \cmidrule(r){16-22} \cmidrule(r){23-29} & \(C_{0}\)& \(C_{1}\) & \(C_{2}\) & \(C_{3}\) & \(C_{4}\) & \(C_{5}\) & Avg. & \(C_{0}\) & \(C_{1}\) & \(C_{2}\) & \(C_{3}\) & \(C_{4}\) & \(C_{5}\) & Avg. & \(C_{0}\)& \(C_{1}\) & \(C_{2}\) & \(C_{3}\) & \(C_{4}\) & \(C_{5}\)& Avg.  & \(C_{0}\)& \(C_{1}\) & \(C_{2}\) & \(C_{3}\) & \(C_{4}\) & \(C_{5}\)& Avg.  \\ \midrule   
LCNN \cite{asvspoof2021} & 8.70 & 8.22 & 8.13 & 9.32 & 8.67 & 10.30 & 8.89 & 3.85 & 4.31 & 4.25 & 4.49 & 4.67 & 5.47 & 4.51 & 2.05 & 2.32 & 2.32 & 2.32 & 2.63 & 3.22 & 2.47 & 0.90 & 1.38 & 1.36 & 1.46 & 1.76 & 2.17 & 1.50\\
RawNet2 \cite{rawnet2} & 23.40 & 20.04 & 20.13 & 20.59 & 21.22 & 22.88 & 21.38 & 4.80 & 3.44 & 3.57 & 4.01 & 4.47 & 6.28 & 4.43 & 2.35 & 1.82 & 1.84 & 2.17 & 2.47 & 3.93 & 2.43 & 2.05 & 1.33 & 1.32 & 1.39 & 1.61 & 2.51 & 1.70\\
AASIST \cite{aasist} & 8.15 & 4.78 & 4.82 & 5.05 & 5.58 & 6.92 & 5.88 & 6.30 & 4.03 & 4.07 & 4.22 & 4.41 & 5.34 & 4.73 & 2.55 & 1.65 & 1.65 & 1.78 & 2.01 & 2.96 & 2.10 & 0.90 & 0.81 & 0.83 & 0.88 & 0.97 & 1.54 & 0.99\\
AASIST-L \cite{aasist} & 9.60 & 8.28 & 8.29 & 8.76 & 9.46 & 11.43 & 9.30 & 8.55 & 5.45 & 5.45 & 5.65 & 5.97 & 7.58 & 6.44 & 3.95 & 3.33 & 3.33 & 3.52 & 3.93 & 5.07 & 3.86 & 2.70 & 2.41 & 2.41 & 2.57 & 2.78 & 3.80 & 2.78\\
RawGAT-ST \cite{rawgat} & 5.60 & 3.77 & 3.81 & 4.05 & 4.42 & 5.44 & 4.52 & 3.15 & 2.22 & 2.28 & 2.39 & 2.67 & 3.69 & 2.74 & 2.80 & 1.31 & 1.31 & 1.41 & 1.53 & 2.13 & 1.75 & 1.75 & 0.60 & 0.63 & 0.71 & 0.85 & 1.60 & 1.02 \\
FC-LFCC \cite{eusipcohaohan} & 13.44 & 14.93 & 14.99 & 15.59 & 16.23 & 18.26 & 15.58 & 9.39 & 9.87 & 9.91 & 10.43 & 11.25 & 13.49 & 10.72 & 6.03 & 7.01 & 7.06 & 7.62 & 8.40 & 10.34 & 7.74 & 4.98 & 5.35 & 5.44 & 5.93 & 6.56 & 8.29 & 6.09\\
MGAA-LFCC \cite{nipshaohan} & 7.25 & 7.69 & 7.72 & 8.12 & 8.90 & 11.13 & 8.47 & 3.80 & 3.78 & 3.82 & 4.11 & 4.73 & 6.20 & 4.41 & 2.00 & 2.23 & 2.32 & 2.59 & 3.30 & 5.10 & 2.92 & 1.50 & 1.78 & 1.76 & 2.10 & 2.44 & 3.94 & 2.25\\
MGAA-CQCC \cite{nipshaohan} & 9.65 & 10.14 & 10.27 & 10.75 & 11.43 & 13.73 & 10.99 & 5.98 & 6.13 & 6.21 & 6.57 & 7.15 & 9.80 & 6.97 & 3.73 & 4.08 & 4.05 & 4.48 & 5.18 & 7.50 & 4.84 & 3.20 & 3.15 & 3.14 & 3.45 & 4.18 & 6.44 & 3.93\\
MGAA-MFCC \cite{nipshaohan} & 4.05 & 4.86 & 4.95 & 5.28 & 5.80 & 7.58 & 5.44 & 2.62 & 2.37 & 2.41 & 2.57 & 2.98 & 4.33 & 2.88 & 1.85 & 1.28 & 1.28 & 1.46 & 1.78 & 2.52 & 1.70 & 0.85 & 0.75 & 0.77 & 0.84 & 1.04 & 1.71 & 0.99\\
\midrule
S-MGAA-LFCC & {4.58} & {4.57} & {4.66} & {4.95} & {5.66} & {7.51} & {5.33} & {1.86} & {1.93} & {2.00} & {2.26} & {2.72} & {3.97} & {2.46} & {0.98} & {0.94} & {0.95} & {1.13} & {1.49} & {2.48} & {1.33} & {0.51} & {0.41} & {0.43} & {0.51} & {0.70} & {1.38} & {0.66}\\
S-MGAA-CQCC & {7.07} & {7.10} & {7.16} & {7.51} & {8.16} & {10.21} & {7.87} & {4.03} & {3.64} & {3.68} & {4.04} & {4.68} & {6.83} & {4.48} & {2.13} & {1.96} & {1.96} & {2.21} & {2.75} & {4.50} & {2.59} & {1.11} & {1.07} & {1.07} & {1.28} & {1.59} & {3.14} & {1.54}\\
S-MGAA-MFCC & \textbf{{2.70}} & \textbf{{3.00}} & \textbf{{3.04}} & \textbf{{3.29}} & \textbf{{3.68}} & \textbf{{4.95}} & \textbf{{3.44}} & \textbf{{1.25}} & \textbf{{1.18}} & \textbf{{1.20}} & \textbf{{1.31}} & \textbf{{1.57}} & \textbf{{2.46}} & \textbf{{1.50}} & \textbf{{0.53}} & \textbf{{0.59}} & \textbf{{0.59}} & \textbf{{0.64}} & \textbf{{0.84}} & \textbf{{1.35}} & \textbf{{0.75}} & \textbf{{0.24}} & \textbf{{0.25}} & \textbf{{0.25}} & \textbf{0.28} & \textbf{{0.36}} & \textbf{0.78} & \textbf{0.36}\\
\bottomrule
\end{tabular}}
\vspace{-0.5cm}
\end{table*}

\vspace{-0.2cm}
\subsection{Frequency Compensation Enhanced Module (FCEM)}
\vspace{-0.1cm}

Ultra-short audio inputs lack sufficient temporal features, limiting the model's ability to capture discriminative temporal patterns. 
To mitigate this, we propose the FCEM, which utilizes frequency-domain features for compensation via Multi-scale Frequency Analysis (MFA) and Adaptive Frequency-temporal Interaction (AFI). The FCEM is computed as:
\begin{equation} 
\text{FCEM}(\delta) = \mathcal{F}({\mathcal{B}_i(\delta)_{i=1}^{3}}, {\mathcal{G}_j(\delta)_{j=1}^{3}}) \odot \mathcal{A}(\delta),
\end{equation}
where $\mathcal{B}_i(\cdot)$ denotes the multi-scale frequency branches, $\mathcal{G}_{j}(\cdot)$ the adaptive pooling operations, $\mathcal{F}(\cdot)$ the fusion operation, and $\mathcal{A}(\cdot)$ the frequency-time attention mechanism.

The MFA is conducted through three parallel branches $\mathcal{B}_i(\cdot)$ with varying receptive fields, along with three complementing adaptive frequency pooling operations $\mathcal{G}_{j}(\cdot)$ to capture global frequency patterns, formulated as:
\begin{align}
\mathcal{B}_i(\delta) &= \sigma_{g}(\mathcal{N}_{b}(W^i_{\kappa_2,k_i \times 1}(\delta))), \quad i \in \{1,2,3\},\\
\mathcal{G}_j(\delta) &= 
\begin{cases} 
\mathcal{G}_{max_{j}}(F_{j},T)(\delta), \quad j \in \{1,2\},\\
\mathcal{G}_{avg}(F_{j},T)(\delta), \quad j =3,
\end{cases}
\end{align}
where $W^i_{\kappa_2,k_i \times 1}$ denotes the channel compression convolution operation with $\kappa_2$ set to 2, $(k_i,1)$ the kernel size and $k_i \in \{20, 15, 10\}$, $\mathcal{G}_{max}$ the max pooling, $F_1, F_2, F_3$ the output feature size after pooling operation, with $F_1=F_3=20$, $F_2=30$. This preserves the details of different frequency branches and constructs rich frequency representations.
Then the outputs of $\mathcal{B}_i(\cdot)$ and $\mathcal{G}_{j}(\cdot)$ are resized to match the original dimensions, concatenated, and fused by $\mathcal{F}(\cdot)$:
\begin{equation} 
\mathcal{F}(\delta) = \sigma_{g}(\mathcal{N}_b(\mathcal{V}_c(concat[\mathcal{B}_i(\delta),\mathcal{G}_j(\delta)])).
\end{equation}

The AFI is to effectively capture the most important frequency features across the time dimension, computed by $\mathcal{A}(\delta) = \sigma_{s}(\mathcal{H}_{DW_f}(\delta))$, where $\mathcal{H}_{DW_{f}}$ is depthwise convolution with kernel size $(7,1)$ along the frequency dimension.



\vspace{-0.2cm}
\section{Experiments}

\vspace{-0.2cm}
\subsection{Dataset and Metrics}
\vspace{-0.1cm}

The training dataset was constructed from six publicly available corpora, Fake-or-Real \cite{FoR}, Wavefake \cite{wavefake}, LJSpeech \cite{ljspeech}, MLAAD-EN \cite{mlaad}, M-AILABS \cite{mailabs}, and ASVspoof2021 Logical Access \cite{asvspoof2021}.
Data preprocessing and augmentation strictly followed the protocol in \cite{eusipcohaohan}.
The resulting dataset, denoted as \(\mathcal{D}_{com}\), comprised 640,205 real and 1,191,865 fake utterances, including 30 types of real-world communication degradations \cite{eusipcohaohan,nipshaohan}. 

The ADD-C test dataset \cite{eusipcohaohan} was selected to evaluate the model's performance on real-world communication perturbed ultra-short duration inputs, which included six conditions \(C_0\) to \(C_5\). \(C_0\) corresponds to the clean audio, while \(C_1\) to \(C_5\) represent increasing degradation levels via six codec compressions under five packet loss rates. 
Equal Error Rate (EER) \cite{asvspoof2019,asvspoof2021,wang2024asvspoof} was selected as the evaluation metric to reflect detection performance; lower EER suggests better performance.

\vspace{-0.2cm}
\subsection{Implementation Details and Baselines}
\vspace{-0.1cm}

\(\mathcal{D}_{com}\) was split into training and validation with an 80\%:20\% ratio. Training was conducted with a batch size of 256 for up to 5 epochs, using early stopping \cite{earlystop} with a patience of 3 to prevent overfitting, and utilizing the Cross-Entropy Loss function. The AdamW optimizer \cite{adamw} was used with a cosine annealing scheduler \cite{cos} to dynamically adjust the learning rate.
Four ultra-short audio durations of 0.5s, 1s, 1.5s, and 2s were considered, resulting in temporal dimensions (\(T\)) corresponding to 16, 32, 47, and 63, respectively, with the frequency dimension (\(F\)) remaining 60 in all cases after the FEM module \cite{nipshaohan}.

To ensure fair and rigorous comparison, we reimplemented and evaluated nine SOTA baselines under identical training and evaluation conditions. These include LCNN \cite{asvspoof2021}, RawNet2 \cite{rawnet2}, AASIST \cite{aasist}, AASIST-L \cite{aasist}, RawGAT-ST \cite{rawgat}, FC-LFCC \cite{eusipcohaohan}, and MGAA \cite{nipshaohan}. All models were trained on the same dataset \(\mathcal{D}_{com}\) and evaluated on ADD-C, using an Intel Core i7-12700K CPU and an NVIDIA RTX 3090 GPU (24GB). Hyperparameters followed the configurations specified in the respective referenced literature.




\vspace{-0.3cm}
\subsection{Experimental Results}
\vspace{-0.1cm}

\begin{figure}[ht]
\centering
\includegraphics[width=0.4\textwidth]{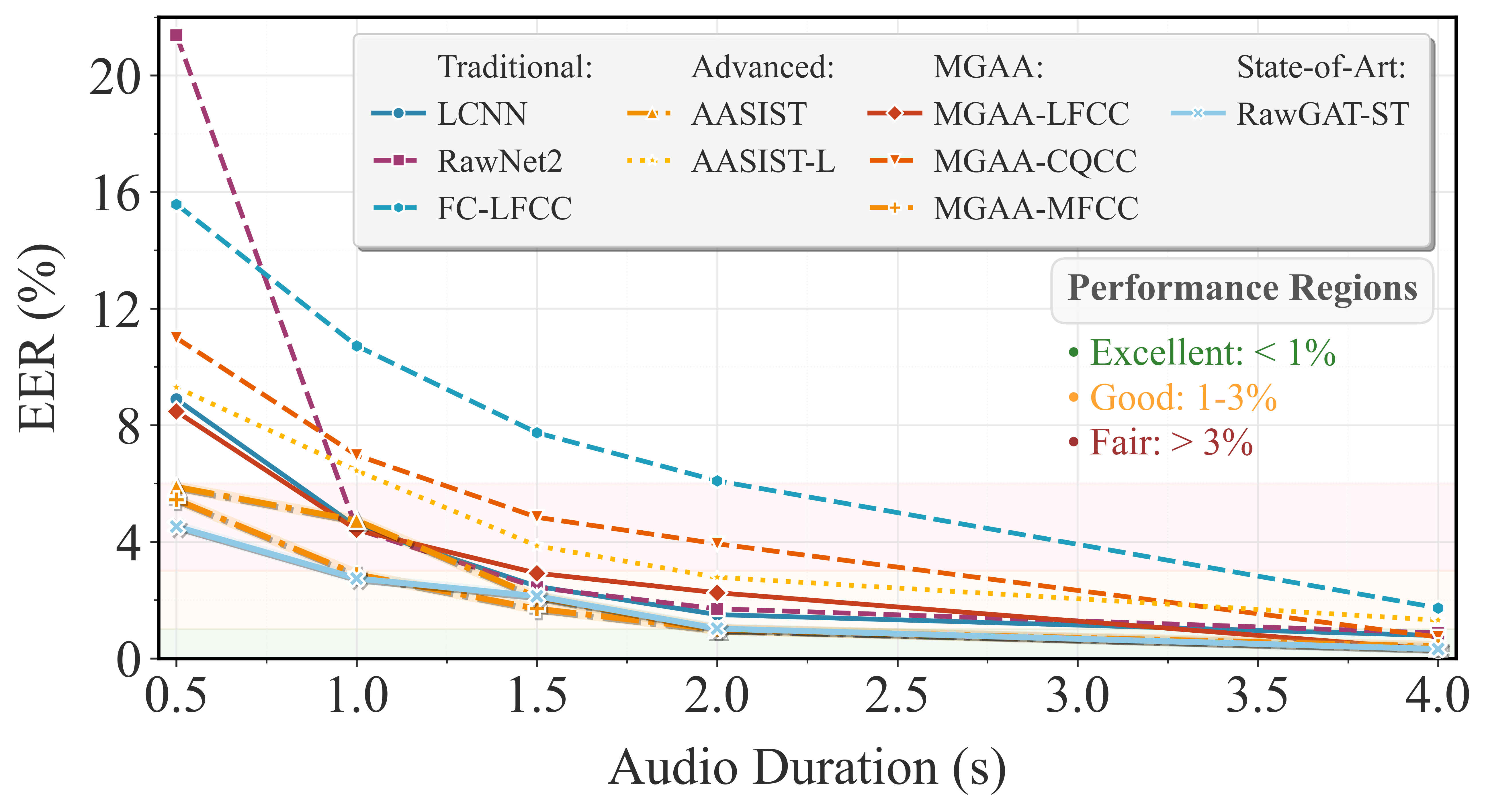}
\vspace{-0.55cm}
\caption{Average EER (\%) of baselines across the $C_0$–$C_5$ conditions for audio durations of 0.5s, 1s, 1.5s, 2s and 4s.}
\label{duration}
\vspace{-0.5cm}
\end{figure}

Table \ref{mainresult} presents the detection performance of nine baselines and the proposed S-MGAA across input durations of 0.5s–2.0s under the conditions: $C_0$–$C_5$. 
Unsurprisingly, longer audio inputs generally improve performance for all methods, since extended durations provide richer temporal and acoustic cues.
However, Fig. \ref{duration} highlights a critical limitation of the nine baseline methods, which experience a notable degradation in performance as duration decreases, particularly in the ultra-short range (0.5–2.0s). This sharp decline underscores the vulnerability of current approaches when operating with severely limited temporal information.

In contrast, as detailed in Table \ref{mainresult}, the proposed S-MGAA achieves consistent performance gains across all durations and conditions. Its advantage is most evident at extremely short inputs. For example, under the challenging 0.5s setting, S-MGAA-MFCC reduces the average EER by 23.89\% compared with the best-performing baseline RawGAT-ST. These results demonstrate that the proposed framework effectively mitigates the information scarcity inherent to ultra-short inputs.

Further evidence is provided in Table \ref{improve}, which compares MGAA and S-MGAA across different features and durations. S-MGAA consistently surpasses MGAA, with relative EER reductions ranging from 28.39\% to 70.67\%, indicating substantial robustness gains from the proposed modules. 
Among input features, the proposed S-MGAA achieved an average improvement of 51.60\% (LFCC), 42.85\% (CQCC), and 51.55\% (MFCC) across 0.5-2.0s, respectively.
Importantly, the improvements across all three feature types demonstrate the generalizability of S-MGAA in extracting discriminative features from severely limited audio duration, allowing reliable detection under real-world communication degradations within the first second of interaction in communication systems.

\begin{table}[ht]
\centering
\vspace{-0.5cm}
\caption{S-MGAA's average performance improvement of EER (\%) compared to MGAA across different audio durations (0.5–2.0s) and input features (LFCC, CQCC, MFCC).
\label{improve}}
\setlength{\tabcolsep}{2.2mm} 
\renewcommand{\arraystretch}{0.7}
\scalebox{0.75}{\begin{tabular}{cccccc}
\toprule
\multirow{2}{*}{Input feature} & \multicolumn{5}{c}{Duration(s)} \\\cmidrule(r){2-6} & 0.5 & 1.0 & 1.5 & 2.0 & Avg. \\ \midrule
LFCC & +37.07\% & +44.22\% & +54.45\% & +70.67\% & +51.60\%\\ 
CQCC & +28.39\% & +35.72\% & +46.49\% & +60.81\%  & +42.85\%\\
MFCC & +36.76\% & +47.92\%  & +55.88\%  & +63.64\%  & +51.55\%\\
\bottomrule
\end{tabular}}
\vspace{-0.4cm}
\end{table}

\vspace{-0.2cm}
\subsection{Ablation Study}
\vspace{-0.1cm}
Table \ref{ablation} shows the impact of removing or altering individual components of our framework.
``Deep'' and ``Shallow'' indicate S-MGAA is
placed only in shallow or deep layers.
Removing either PCEM or FCEM leads to consistent performance degradation across all feature types and durations.
Moreover, placing S-MGAA only in deep or shallow layers underperforms the hybrid design, indicating that distributing modules across different depths enables more effective feature refinement, balancing low-level and high-level representation learning. 
The full S-MGAA achieves the best results, validating the effectiveness of each component and their joint integration.

\begin{table}[ht]
\centering
\caption{Ablation studies of framework components, where \(\checkmark\) and \(\times\) denote inclusion and exclusion, respectively. Results averaged
over three runs.}
\label{ablation}
\setlength{\tabcolsep}{0.7mm}
\scalebox{0.58}{
\begin{tabular}{ccccccccccccccccc}
\toprule
\multirow{3}{*}{MGAA} & \multirow{3}{*}{PCEM} & \multirow{3}{*}{FCEM} & \multirow{3}{*}{Deep} & \multirow{3}{*}{Shallow} & \multicolumn{12}{c}{Avg. EER (\%) \(\downarrow\)} \\\cmidrule(r){6-17} &&&&&\multicolumn{4}{c}{LFCC} & \multicolumn{4}{c}{CQCC} & \multicolumn{4}{c}{MFCC} \\ \cmidrule(r){6-9} \cmidrule(r){10-13} \cmidrule(r){14-17} &&&&& 0.5s & 1.0s & 1.5s & 2.0s& 0.5s & 1.0s & 1.5s & 2.0s& 0.5s & 1.0s & 1.5s & 2.0s \\
\midrule   
\(\times\) & \(\checkmark\) & \(\checkmark\) & \(\checkmark\) & \(\checkmark\) & 5.41 & 2.61 & 1.42 & 0.72 & 8.16 & 4.71 & 2.77 & 1.69 & 3.65 & 1.55 & 0.87 & 0.39 \\

\(\checkmark\) & \(\times\) & \(\checkmark\) & \(\checkmark\) & \(\checkmark\) & 5.84 & 2.80 & 1.49 & 0.80 & 8.30 & 4.79 & 2.91 & 1.78 & 3.64 & 1.73 & 0.86 & 0.47\\

\(\checkmark\) & \(\checkmark\) & \(\times\) & \(\checkmark\) & \(\checkmark\) & 5.49 & 2.56 & 1.40 & 0.70 & 8.18 & 4.64 & 2.82 & 1.67 & 3.54 & 1.61 & 0.82 & 0.40 \\

\(\checkmark\) & \(\checkmark\) & \(\checkmark\) & \(\times\) & \(\checkmark\) & 5.68 & 2.60 & 1.35 & 0.73 & 8.25 & 4.73 & 2.90 & 1.70 & 3.62 & 1.65 & 0.86 & 0.44\\ 

\(\checkmark\) & \(\checkmark\) & \(\checkmark\) & \(\checkmark\) & \(\times\) & 5.78 & 2.62 & 1.45 & 0.79  & 8.47 & 5.01 & 2.97 & 1.79 & 3.81 & 1.64 & 0.88 & 0.43\\ 

\(\checkmark\) & \(\checkmark\) & \(\checkmark\) & \(\checkmark\) & \(\checkmark\) & \textbf{5.33} & \textbf{2.46} & \textbf{1.33} & \textbf{0.66} & \textbf{7.87} & \textbf{4.48} & \textbf{2.59} & \textbf{1.54} & \textbf{3.44} & \textbf{1.50} & \textbf{0.75} & \textbf{0.36}\\ 
\bottomrule
\end{tabular}}
\vspace{-0.5cm}
\end{table}

\vspace{-0.2cm}
\subsection{Complexity and Deployability Analysis}
\vspace{-0.1cm}

\begin{figure}[ht]
\vspace{-0.2cm}
\centering
\includegraphics[width=0.483\textwidth]{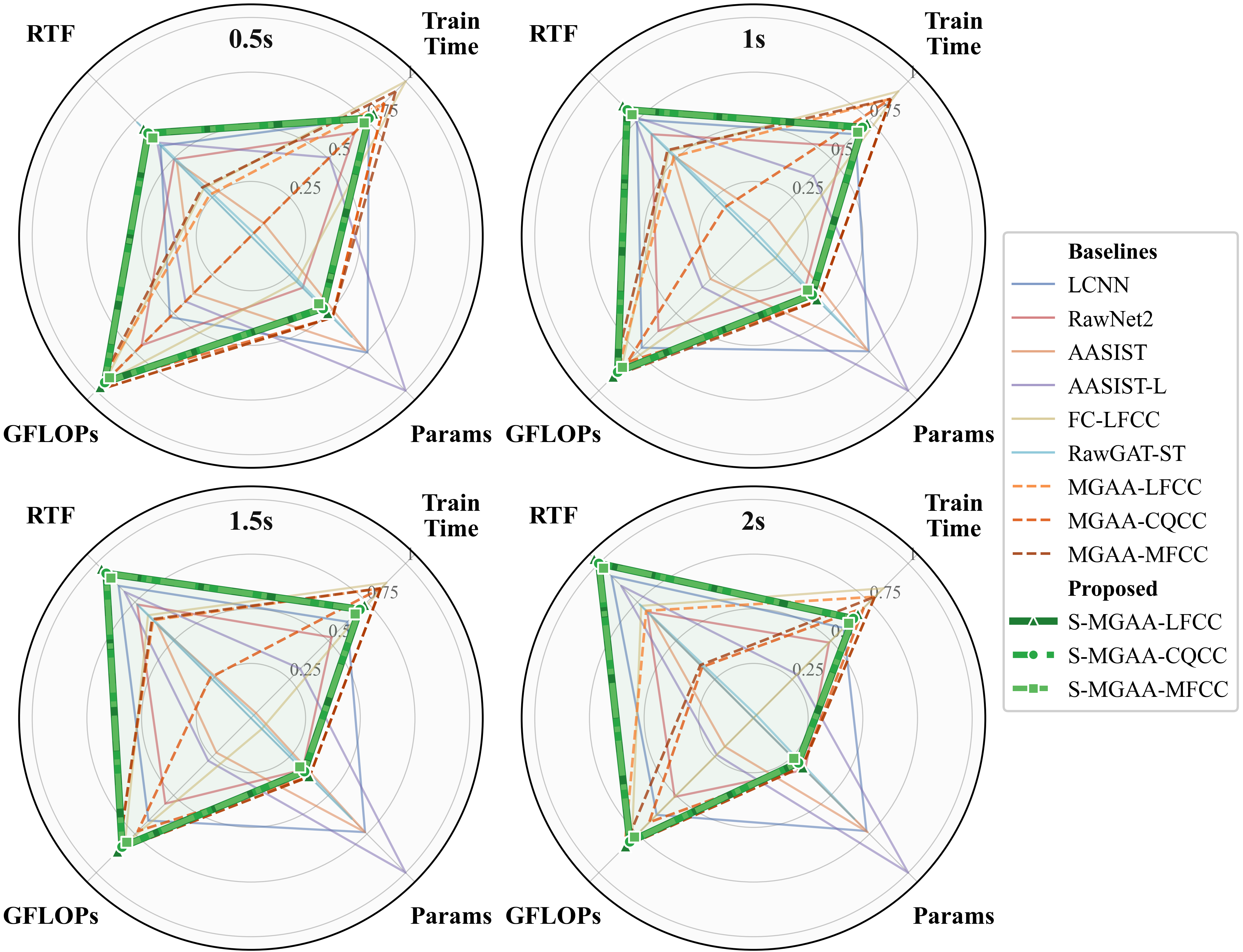}
\vspace{-0.7cm}
\caption{Efficiency comparison across 0.5–2.0s. Radar plots show log‑scaled, per‑duration–normalized metrics (outer position is better after inversion for RTF, training time, GFLOPs, and parameters). Green coloured areas denote the proposed S‑MGAA variants.
}
\label{complex}
\vspace{-0.3cm}
\end{figure}

To assess model complexity and deployability, we compare S‑MGAA (via green coloured) with nine baselines across 0.5-2.0s, as shown in Fig. \ref{complex}.
RawGAT‑ST incurs excessive computational cost (36.12 GFLOPs and 15.78h training time for all durations), AASIST variants require long training time (10.03–17.51h) with multi‑GFLOP requirement (2.23–9.02G).
FC‑LFCC scales poorly in parameters (2.08→7.48M from 0.5→2.0s). MGAA baselines keep GFLOPs low (0.01–0.22G), but RTF fluctuates largely during short inputs (e.g., 1.75-4.82 at 0.5 seconds).
In contrast, S‑MGAA maintains low compute and stable latency across different duration inputs, with FLOPs (0.02→0.08G), parameters (0.99→2.14M), RTF (0.38→0.10), and training time (0.25–0.49h) for 0.5–2.0s. This efficiency-accuracy balance highlights the practicality and robustness of S-MGAA for real-world, resource-constrained deployment.

\vspace{-0.4cm}
\section{Conclusion}
\vspace{-0.2cm}
We proposed S-MGAA, a novel lightweight framework for ADD under real-world communication degradations with ultra-short inputs. The framework enhances discriminative representation learning from limited audio inputs, enabling reliable detection within durations as short as a greeting phrase. Experiments across multiple features and degradations demonstrated that S-MGAA consistently outperforms nine SOTA baselines while maintaining low computational cost, marking an important step towards a practical, real-time deployable, and reliable ADD method.
\href{https://github.com/Haohan-SHI/S-MGAA}{Codes are available.}

\bibliographystyle{IEEEbib}
\bibliography{refs}

\end{document}